%% file: ordinalsFO.tex
\documentclass{llncs}

\usepackage{amsmath,gastex}
\usepackage{amssymb}
\usepackage{multirow}
\usepackage{epic,eepic,latexsym}
\usepackage[english]{babel}

\input{macros-ordinalsFO}

\begin{document}
\title{ Tree Automata Make Ordinal Theory Easy  }
\author{Thierry Cachat} 
\institute{LIAFA/CNRS UMR 7089 \& Universit{\'e} Paris 7, France
	\\ email: \texttt{txc@liafa.jussieu.fr}
}

\maketitle

\begin{abstract} 
We give a new simple proof of the decidability of the First Order Theory of 
$(\w^{\w^i},+)$ and the Monadic Second Order Theory of $(\w^i,<)$, improving the 
complexity in both cases. Our algorithm is based on tree automata and a new
representation of (sets of) ordinals by (infinite) trees.
\end{abstract}

%
%
\section{Introduction}
%

The connections between automata and logic have been fruitful for many 
years, see \cite{dagst} for an introduction. In 1960 Büchi \cite{Buechi60}
showed that sets of finite words can be equivalently defined by Monadic
Second Order (MSO) formulas and by finite automata. This gives in particular
a decision procedure for this logic. This result has been extended later
to other classes of structures and automata: MSO over infinite words and Büchi
automata in \cite{Buechi62}, MSO over transfinite ordinals and transfinite 
automata \cite{Buechi65a}, MSO over the full binary tree and Rabin automata in 
\cite{Rabin69}, MSO over graphs of the Caucal hierarchy and graph automata
\cite{icalp03,KupfermanVardi00}.

The decidability of the first order logic over the integers with addition, also known 
as Presburger arithmetic, can be easily obtained by using finite automata reading
binary representation of numbers. A central idea in all these results 
is that formulas can be represented by automata: by 
induction on the formula one can build an automaton accepting exactly the 
models of the formula. See \cite{Weyer01} 
for a clear exposition of many of the previous results.

More recently many authors have used automata to improve the complexity
of certain decisions procedures. In particular in \cite{Klaedtke04} the
Presburger arithmetic is considered and in \cite{Maurin96} the first order
theory of the ordinals with addition.

We address in this article the decision algorithms for the First Order theory
(FO) of $(\w^{\w^i},+)$ and the Monadic Second Order theory (MSO) of
$(\w^i,<)$ for any integer $i$. Our proposal is to use finite labeled trees to 
represent ordinals and infinite trees to represent sets of ordinals. Then one can
use tree automata to represent formulas (namely, all their models).
In this way we improve the best known complexity, and we hope that our constructions 
are easier to understand than previous ones. Note that already $\MSO(\w,+)$ is 
undecidable, and the decision procedure for $\MSO(\w,<)$ has a non 
elementary lower bound. 
In \cite{Dershowitz93} trees are already used to represent ordinals, but only 
termination of preocesses is considered.
Our infinite trees in \Sec{sec:MSO} are close to those in \cite{BruyCartonSeniz03}, 
where only inclusion of languages is considered.

The paper is organized as follows. The next section is concerned with the
first order theory. After recalling definitions we present our tree encoding
and our decidability proof. In \Sec{sec:MSO} the encoding is adapted to the
Monadic Second Order theory, before comparisons to other results
and techniques are given.

%
%
\section{Decidability of the First Order Theory of $(\w^\w,+)$}
%
\label{sec:FO}

\subsection{Definitions: Ordinal Addition, First Order Logic, Tree Automata}
%

We assume basic knowledge about ordinals, see e.g. 
\cite{Rosenstein82,Sierpinski65}.
An {\em ordinal} is a well and totally ordered set.
It is either $0$ or a successor ordinal of the form $\beta+1$ or a limit ordinal.
The first limit ordinal is denoted $\w$. For all ordinal $\a$:
$\b<\a \Equ \b\in \a$ and $\a=\{\b : \b<a\}$. The set of natural numbers is identified
with $\w$.
Recall e.g. that $1+\w=\w=2\w$ and $\w+\w^2=\w^2$ but $\w+1\neq \w \neq \w 2$.
By the Cantor Normal Form theorem, for all $0< \a< \w^\w$ there exist
unique integers $p,n_0,n_1,\dots,n_p$ such that $n_p>0$ and
\debeqno
	\a=\w^p n_p+\w^{p-1}n_{p-1}+\dots+  \w^{1}n_1+n_0 \ .
\fineqno
Ordinal addition has an {\em absorption} property: for any $p<p'$, $\w^p+\w^{p'}=\w^{p'}$.
Given two ordinals $\a=\w^p n_p+\dots+  \w^{1}n_1+n_0$ and
$\a'=\w^{p'} n'_{p'}+\dots+  \w^{1}n'_1+n'_0$ both written in Cantor 
Normal Form, the ordinal $\a+\a'$ is
\debeqno
	\w^p n_p+\dots+\w^{p'} (n_{p'}+n'_{p'})+\dots+  \w^{1}n'_1+n'_0\ .
\fineqno
Formulas of the {\em First Order Logic} (FO) over $(\w^\w,+)$ are built from 
\debitem
\item a countable set of individual variables $x,y,z,\dots$
\item the addition $+$, seen as a ternary relation, 
\item the Boolean connectives $\neg$, $\wedge$, $\vee$, $\imp$ and $\equ$,
\item first order quantification $\exists$ over individual variables ($\forall$ is seen
as an abbreviation of $\neg\exists\neg$).
\finitem
\debexm \label{exm:1}
The order relation $x\leq y$ can be easily defined as $\exists z: x+z=y$.	\\
The relation $x<y$ is defined by $\neg (y\leq x)$.	\\
The ordinal $0$ is the only ordinal $x$ such that $\neg\exists y: y<x$ or
equivalently such that $x+x=x$. \\
The equality between $x$ and $y$ can be defined e.g. by $x\leq y \wedge y\leq x$. \\
The ordinal $1$ is definable by $\phi(x)=(x>0)\wedge \neg\exists y (0<y \wedge y<x)$.
\finexm
\debexm \label{exm:2}
The first limit ordinal, $\w$, is the only ordinal satisfying the formula 
\debeqno
	\varphi_{1}(x) & = & (x>0)\wedge \forall y (y<x \imp y+1<x) \wedge \\
			& & \forall x'[ (x'>0)\wedge \forall y (y<x' \imp y+1<x')\ \imp\ x\leq x'] \ .
\fineqno
Similarly and by induction $\w^{i+1}$ is defined by
\debeqno
	\varphi_{i+1}(x) & = & (x>0)\wedge \forall y (y<x \imp y+\w^i<x) \wedge \\
			& & \forall x'[ (x'>0)\wedge \forall y (y<x' \imp y+\w^i<x')\ \imp\ x\leq x'] \ .
\fineqno
\finexm
A {\em finite binary tree} $T$ is a finite prefix closed subset of $\{a,b\}^*$. 
The root is the empty word $\eps$, and for all $u\in\{a,b\}^*$, $u a$ is the left successor of $u$ and 
$u b$ the right one. For simplicity we impose that each node has $0$ or $2$ 
successors: $\forall u\in\{a,b\}^*$, $u a\in T \Equ u b\in T$. A leaf has no successor.
Given a finite alphabet $\S$, a $\S$-labeled tree is a couple 
$\lc T,\lambda\rc$ where $T$ is a tree and $\lambda$ is a function  
$\lambda:T\vers \S$.
A {\em tree automaton} is a tuple $(Q,\S,\D,I,F)$ where $Q$ is a finite
set of states, $\S$ is a finite alphabet, $\D\inc Q\ti \S\ti Q\ti Q$ is the transition relation,
$I\inc Q$ and $F\inc Q$ are the sets of initial and accepting states (``final states''). 
A $\S$-labeled tree is 
accepted by such a tree automaton iff there exists a run  $\rho:T\vers Q$ such that
\debeqno
	\rho(\eps)\in F\mbox{, and }\forall u\in T:
		& \mbox{ either } & (\rho(u),\lambda(u),\rho(ua),\rho(ub))\in \D\\
		& \mbox{ or } & u \mbox{ is a leaf }(u a\not\in T) \mbox{ and }\rho(u)\in I \ .
\fineqno
This presentation is unusual: the labels at the leafs are not important in our 
constructions.
These (bottom up) tree automata can be determinized by a usual subset construction.
By exchanging initial and final states they can be seen as top down 
automata.

\subsection{Binary Trees Representing Ordinals}
%

Ordinals less than $\w^\w$ can be easily represented by finite binary trees.
The tree representing $\a=\w^p n_p+\dots+  \w^{1}n_1+n_0$ (where $n_p>0$)
has a leftmost branch of length (at least) $p$. At depth $i$ on 
this branch a right branch is attached, holding the binary encoding 
of the number $n_i$. For example the
ordinal $\w^3.5+\w.3+8$ is represented essentially as the following tree.
\\[.5 em]
{ \unitlength=0.7mm
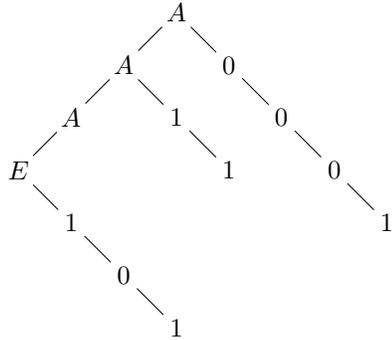
\begin{picture}(75,65)(-33,-62)
	\gasset{Nadjust=wh,Nadjustdist=2,Nframe=n}
	\node(r)(0,0){$A$}
	\node(1)(10,-10){$0$}
	\node(11)(20,-20){$0$}
	\node(111)(30,-30){$0$}
	\node(1111)(40,-40){$1$}
	
	\node(0)(-10,-10){$A$}
	\node(01)(0,-20){$1$}
	\node(011)(10,-30){$1$}
	
	\node(00)(-20,-20){$A$}

	\node(000)(-30,-30){$E$}
	\node(0001)(-20,-40){$1$}
	\node(00011)(-10,-50){$0$}
	\node(000111)(0,-60){$1$}

	\gasset{AHnb=0}
	\drawedge(r,0){$$}
	\drawedge(0,00){$$}
	\drawedge(00,000){$$}
	
	\drawedge(r,1){$$}
	\drawedge(1,11){$$}
	\drawedge(11,111){$$}
	\drawedge(111,1111){$$}
	
	\drawedge(0,01){$$}
	\drawedge(01,011){$$}
	\drawedge(000,0001){$$}
	\drawedge(0001,00011){$$}
	\drawedge(00011,000111){$$}
\end{picture} }
\ 
\begin{minipage}[b]{6.5 cm}
The letter $E$ marks the last position where there is a non zero right branch.
We allow {\em all} possible ways to add
dummy symbols $\#$ at the bottom of the tree. There are not represented on the 
picture, but they are needed for every node to have 0 or 2 successors (not 1). 
To be more formal the set of tree representations of a given ordinal 
$\a=\w^p n_p+\dots+  \w^{1}n_1+n_0$ is exactly the language accepted by the tree 
automaton to be defined next. 
The initial state is $q_\#$, the accepting state $q_0$.
\end{minipage}
\\
If $\s_i^0 \s_i^1 \dots \s_i^{m_i}$ is the (little endian) binary encoding of 
$n_i$: $n_i=\sum_{j=0}^{m_i}  2^j \s_i^j$, then the transitions are:
\debeqno
	(q_i,A,q_{i+1},p_i^0) \mbox{ if } i<p \mbox{ and } n_i>0 & \qquad & 
			(p_i^j,\s_i^j,q_\#,p_i^{j+1}) \mbox{ if } j<m_i \\
	(q_i,A,q_{i+1},q_\#) \mbox{ if } i<p \mbox{ and } n_i=0 & \qquad & 
		(p_i^j,\s_i^j,q_\#,q_\#) \mbox{ if } j=m_i \\
	(q_i,E,q_\#,p_i^0)  \mbox{ if }  i=p & \qquad &
		(q_\#,\#,q_\#,q_\#)
\fineqno
In the special case where $\a=0$ we have a transition $(q_0,\#,q_\#,q_\#)$. 
We denote $T_\a$ the tree coding an ordinal $\a$.
\subsection{Decidability Using Tree-Automata}
%

We adapt a well known method for proving decidability of logic theories.
A single tree over the alphabet $\{A,E,\#,0,1\}^k$ represents the values of
$k$ variables by superposing $k$ corresponding trees (and adding dummy
symbols $\#$).
For every formula $\p\in \FO(\w^\w,+)$ with free variables $x_1,\dots,x_k$ we want
to build a tree automaton over the alphabet $\{A,E,\#,0,1\}^k$ such that a tree 
is  accepted by this automaton iff the corresponding valuation of the variables 
satisfies  $\p$. 
This can be done by induction on the formula. The case of Boolean connectives 
is easy using standard automata techniques of product and complementation, see
\cite{tata97}. Existential quantification results in projecting out the corresponding
variable. The main point is to define an automaton recognizing the relation
$x+y=z$, and this is easy with our coding.

In the following transitions $\tr{\#}{1}{0}$ represents a letter from $\{A,E,\#,0,1\}^3$
where the first component is $\#$, the second is $1$ and the third is $0$. 
These components are letters from $T_x$, $T_y$ and $T_z$ respectively.
The symbols $\s,\d$ represent digits from $\{0,1\}$ and $*$ represents any letter. 
The accepting state is $r$. 
Because of the absorption property, above symbol $E$ of $T_y$, trees $T_y$ and $T_z$ 
must coincide.
State $q_y$ checks that $T_y$ and $T_z$ coincide on the
corresponding right branch. Similarly $q_x$ checks that $T_x$ and $T_z$
coincide.  
State $r_y$ checks that $T_y$ and $T_z$ coincide on the rest of the tree. 
Similarly $r_x$ checks that $T_x$ and $T_z$ coincide. 
\debeqno
	(r,\tr{\#}{\#}{\#},q_\#,q_\#) & \qquad  (q_\#,\tr{\#}{\#}{\#},q_\#,q_\#) & \\
	(r,\tr{A}{A}{A},r,q_y) & \qquad (q_y,\tr{*}{\s}{\s},q_\#,q_y) & \qquad 
		(q_y,\tr{*}{\#}{\#},q_\#,q_y) \qquad (q_y,\tr{\#}{\#}{\#},q_\#,q_\#) \\
	(r,\tr{E}{A}{A},r_y,q_y) & \qquad  (r_y,\tr{\#}{A}{A},r_y,q_y) & \qquad  (r_y,\tr{\#}{E}{E},q_\#,q_y) \\
	(r,\tr{A}{E}{A},r_x,q_0) & \qquad  (r_x,\tr{A}{\#}{A},r_x,q_x) & \qquad 
		(r_x,\tr{E}{\#}{E},q_\#,q_x) \\
	(r,\tr{E}{E}{E},q_\#,q_0) & \qquad  (q_x,\tr{\s}{*}{\s},q_\#,q_x) &  \qquad 
		(q_x,\tr{\#}{*}{\#},q_\#,q_x) \qquad (q_x,\tr{\#}{\#}{\#},q_\#,q_\#)
\fineqno
The states $q_0$  and $q_1$ are in charge of the binary addition with carries.
\debeqno
	(q_0,\tr{\s}{\d}{\s\mbox{ XOR } \d},q_\#,q_{\s\mbox{\scriptsize AND } \d})  & \qquad 
		(q_0,\tr{\s}{\#}{\s},q_\#,q_{x}) & \qquad (q_0,\tr{\#}{\s}{\s},q_\#,q_{y})  \\	(q_1,\tr{\s}{\d}{\neg(\s\mbox{ XOR } \d)},q_\#,q_{\s\mbox{\scriptsize OR } \d}) & \qquad
		(q_1,\tr{\s}{\#}{\neg \s},q_\#,q_{\s}) & \qquad (q_1,\tr{\#}{\s}{\neg \s},q_\#,q_{\s})  \\
	(q_0,\tr{\#}{\#}{\#},q_\#,q_\#) & \qquad (q_1,\tr{\#}{\#}{1},q_\#,q_\#) & 
\fineqno
Some details are omitted here for the sake of simplicity. In state $q_y$, after
reading $\#$ on the first component, one should check that only $\#$ appears.
And the most significant bit of each number should be $1$ to have a standard
representation. 
It is left to the reader to add intermediate states to check that the trees $T_x$, $T_y$ and 
$T_z$ are well formed. That is needed when the automata defining $T_x$, $T_y$ or
$T_z$ were obtained by complementation (see below).
Let $\tow$ stand for the ``tower of exponentials'' function, \ie, $\tow(0,n)=n$ and 
$\tow(k+1,n)=2^{\tow(k,n)}$. 
\debthm \label{thm:FO1}
The First Order Theory of $(\w^\w,+)$ is decidable in time \\        
$\bigO(\tow(n,c))$, for some constant $c$, where $n$ is the length of the formula.
\finthm
To our knowledge the best known algorithm for deciding $\FO(\w^\w,+)$ goes via a (linear)
reduction to the Weak Monadic Second Order logic of $(\w^\w,<)$, which in 
turn is decidable in time $\bigO(\tow(6n,c'))$ \cite{Maurin96}. 
See Section~\ref{sec:MSO} for the definition of this logic. 
\debdem
By induction on the formula $\p\in \FO(\w^\w,+)$ one can construct a tree automaton
$\ar_\p$ accepting exactly all valuations satisfying $\p$. A valuation is here a tree
labeled over $\{A,E,\#,0,1\}^k$, where $k$ is the number of free variables in $\p$.
\debitem
\item If $\p$ is an atomic proposition, it is of the form $x+y=z$ and we have seen how
	to construct $\ar_\p$.
\item If $\p$ is of the form $\neg \p'$, by induction $\ar_{\p'}$ is constructed. We can
	determinize and complement it \cite{tata97}, and intersect with the automaton
	describing the allowed representation of ordinals, to obtain $\ar_\p$. 
\item If $\p$ is of the form $\p_1\wedge \p_2$, by induction $\ar_{\p_1}$ and 
	$\ar_{\p_2}$ are constructed. Rearrange the order of the variables,
	build the product of $\ar_{\p_1}$ and 
	$\ar_{\p_2}$. Declare a state $\lc q_1,q_2\rc$ final iff both $q_1$ and $q_2$
	are final. \cite{tata97}
\item Similarly if $\p$ is of the form $\p_1\vee \p_2$, rearrange the variables, build 
      the product and declare a state $\lc q_1,q_2\rc$ final iff $q_1$ or $q_2$ is final.
\item If $\p$ is of the form $\p_1\imp \p_2$, first determinize $\ar_{\p_1}$ and 
	$\ar_{\p_2}$, then build the product, and declare a state $\lc q_1,q_2\rc$ 
	final iff $(q_1 \in F_1)\Imp (q_2\in F_2)$.
\item Similarly if $\p$ is of the form $\p_1\equ \p_2$, determinize $\ar_{\p_1}$ and 
	$\ar_{\p_2}$, build the product, and declare a state $\lc q_1,q_2\rc$ 
	final iff $(q_1 \in F_1)\Equ (q_2\in F_2)$.
\item If $\p$ is of the form $\exists x \p'$, then the input alphabet of the automaton 
      $\ar_\p'$ is $\{A,E,\#,0,1\}^k$, where $k$ is the number of free variables
      in $\p'$. Project out the component corresponding to the
        variable $x$ to get the automaton $\ar_\p$ that non-deterministically guesses the
	value of $x$.
\finitem
At the end of the procedure it remains to determine whether $\ar_\p$ accepts a tree 
(labeled over an empty alphabet). This can be done in polynomial time by marking the 
states reachable from the initial states.
Note that the cases of conjunction and disjunction does not need determinization.
This is possible with a bottom up tree automaton, where the acceptance condition
is checked only once, at the root.
\\  
Like for many automata based decision procedures, the most expensive step is the 
determinization of automata. It costs exponential time and the result is an automaton 
of exponential space. The number of steps of the construction is the number of
Boolean connectives and quantifiers of the formula, whereas the constant $c$ is
essentially the number of states of the automaton for $x+y=z$.
\findem

To slightly improve the complexity one can easily construct directly automata recognizing
the relations $x=y$, $x<y$, $x\leq y$ of Example~\ref{exm:1}. Of course every ordinal
$\w^i$ can also be easily defined directly, without using the formulas
of Example~\ref{exm:2}.

It is also possible to replace $\imp$ and $\equ$ by equivalent formulas using only
$\neg, \wedge$ and $\vee$ and to push negations symbols inwards (using De 
Morgan's laws, etc). See \cite{Klaedtke04} for a careful discussion about the cost 
of these transformations: they can increase the length of the formula and add new 
quantifiers. Here we do not assume that the formula is in prenex normal form.

%
%
\subsection{Beyond $\w^\w$} \label{sec:beyond}
%

By using a new letter ($B$) in the alphabet, it is possible to encode
ordinals greater than $\w^\w$. 
Any ordinal $\b<\w^{\w^2}$ can be uniquely written in the form
\debeqno
        \w^{\w.p}\a_p+ \dots + \w^{\w.2}\a_2+\w^{\w}\a_1+\a_0\ ,
	\mbox{ where } p<\w, \a_i<\w^\w, \a_p>0\ .
\fineqno
\begin{minipage}[b]{8 cm}
and we can encode it as a tree where each $T_{\a_i}$ appears as a subtree.
Namely the leftmost branch will have length $p$. At depth $i$ 
on this branch the tree $T_{\a_i}$ is attached. The skeleton of
the tree is depicted on the right. It is easy to see that a tree automaton can
recognize the relation $x+y=z$, and that the proof of Theorem~\ref{thm:FO1}
carries over. Note that the letter $B$ is used here only for clarity, one
could use $A$ instead.
\end{minipage}
{ \unitlength=0.8mm
\begin{picture}(50,45)(-30,-42)
	\gasset{Nadjust=wh,Nadjustdist=2,Nframe=n}
	\node(r)(0,0){$B$}
	\node(0)(-10,-10){$B$}
	\node(00)(-20,-20){$$}
	\node(001)(-10,-30){$$}
	\node(000)(-25,-25){$E$}

	\gasset{Nadjust=wh,Nadjustdist=3,Nframe=y}
	\node[dash={0.7 1.5}0](1)(10,-10){$T_{\a_0}$}		
	\node[dash={0.7 1.5}0](01)(0,-20){$T_{\a_1}$}
	\node[dash={0.7 1.5}0](0001)(-15,-35){$T_{\a_p}$}

	\gasset{AHnb=0}
	\drawedge(r,0){$$}
	\drawedge[dash={0.5 1}0](0,000){$$}
	\drawedge(r,1){$$}
	\drawedge(0,01){$$}
	\drawedge(000,0001){$$}
\end{picture}}

This can be generalized by induction, and for all $i<\w$
we can encode ordinals less than $\w^{\w^i}$.
\debthm \label{thm:FO2}
     For each $i<\w$ there exists a constant $c_i$ such that the 
     First Order Theory of $(\w^{\w^i},+)$ is decidable in time 
     $\bigO(\tow(n,c_i))$, where $n$ is the length of the formula.
\finthm
Note that the height of the tower of exponentials do not depend on $i$, and that
$c_i$ is linear in $i$. When considering $\FO(\w^{\w^i},+)$, even the
ordinal $1$ is coded by a tree of depth at least $i$: we need each tree to
have the same skeleton to allow the automaton to proceed the addition
locally. It was already noticed (without proof) in \cite{Delhomme04} that any
ordinal $\a<\w^{\w^\w}$ is tree-automatic, that is to say that the structure
$(\a,<)$ ---without addition--- is definable using tree-automata. Moreover 
\cite{Delhomme04} proves that any tree-automatic ordinal is less than 
$\w^{\w^\w}$.

%
%
\section{Monadic Second Order Theory of $(\w^k,<)$}
%
\label{sec:MSO}

In this section we use full {\em infinite} binary trees. They are given by a mapping
$\lambda:\{a,b\}^*\vers \S$ for some finite alphabet $\S$. Their domain is
always $\{a,b\}^*$ so we do not need to mention it.
One can adapt the idea of Section~\ref{sec:FO} to represent {\em sets} of ordinals.
Given a subset $S\inc \w^2$ it is represented by the tree $\lambda:\{a,b\}^*\vers \{0,1\}$
such that
\debeqno
	&& \forall i,j\geq 0: \lambda(a^i b^j)\in\{0,1\} 
	        \qquad \forall u\not\in a^*b^*: \lambda(u)=\# \\
        &&        \forall i,j\geq 0:  \lambda(a^i b^j)=1\ \Equ\ \w.i+j\in S \ .
\fineqno
So positions are associated to ordinals according to the left tree of the next picture. 
Accordingly the right tree represents the set $\{0,\w+1,\w+2,\w.2+2,\w.3\}$. In this
way one can represent any subset of $\w^2$.
\\
{ \unitlength=0.8mm
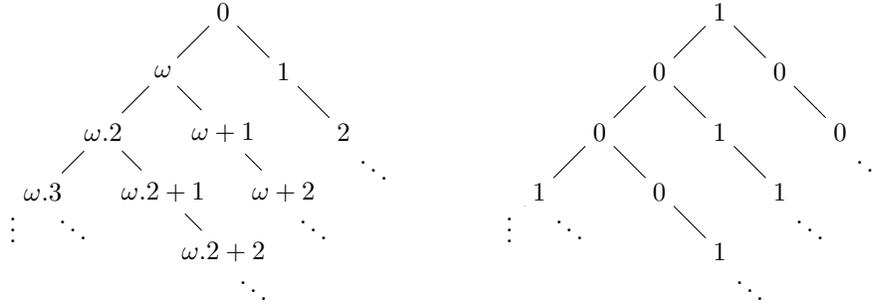
\begin{picture}(70,52)(-40,-49) \label{fig:positions}
        \gasset{Nadjust=wh,Nadjustdist=2,Nframe=n}
	\node(r)(0,0){$0$}
	\node(1)(10,-10){$1$}
	\node(11)(20,-20){$2$}
	\node(111)(25,-25){$\ddots$}		
	\node(0)(-10,-10){$\w$}
	\node(01)(0,-20){$\w+1$}
	\node(011)(10,-30){$\w+2$}
	\node(0111)(15,-35){$\ddots$}
	\node(00)(-20,-20){$\w.2$}
	\node(001)(-10,-30){$\w.2+1$}
	\node(0011)(0,-40){$\w.2+2$}
	\node(00111)(+5,-45){$\ddots$}
	\node(000)(-30,-30){$\w.3$}
	\node(0001)(-25,-35){$\ddots$}
	\node(0000)(-35,-35){$\vdots$}

	\gasset{AHnb=0}
	\drawedge(r,0){$$}
	\drawedge(0,00){$$}
	\drawedge(00,000){$$}
	\drawedge(r,1){$$}
	\drawedge(1,11){$$}
	\drawedge(0,01){$$}
	\drawedge(01,011){$$}
	\drawedge(00,001){$$}
	\drawedge(001,0011){$$}
\end{picture} 
\hf
\begin{picture}(70,52)(-40,-49)
	\gasset{Nadjust=wh,Nadjustdist=2,Nframe=n}
	\node(r)(0,0){$1$}
	\node(1)(10,-10){$0$}
	\node(11)(20,-20){$0$}
	\node(111)(25,-25){$\ddots$}
	\node(0)(-10,-10){$0$}
	\node(01)(0,-20){$1$}
	\node(011)(10,-30){$1$}
	\node(0111)(15,-35){$\ddots$}	
	\node(00)(-20,-20){$0$}
	\node(001)(-10,-30){$0$}
	\node(0011)(0,-40){$1$}
	\node(00111)(+5,-45){$\ddots$}
	\node(000)(-30,-30){$1$}
	\node(0001)(-25,-35){$\ddots$}
	\node(0000)(-35,-35){$\vdots$}

	\gasset{AHnb=0}
	\drawedge(r,0){$$}
	\drawedge(0,00){$$}
	\drawedge(00,000){$$}
	\drawedge(000,0000){$$}	
	\drawedge(r,1){$$}
	\drawedge(1,11){$$}
	\drawedge(0,01){$$}
	\drawedge(01,011){$$}
	\drawedge(00,001){$$}
	\drawedge(001,0011){$$}
\end{picture}  }
Languages of infinite trees can be defined by top down 
{\em Muller automata} \cite{Niessner01}. 
A Muller automaton $\ar$ is a tuple $(Q,\S,\D,I,\fr)$ where $Q,\S,\D$ are the same as
in Section~\ref{sec:FO}, $I\inc Q$ is the set of initial states and 
$\fr\inc \pr(Q)$ is the acceptance component ($\pr(Q)$ is the powerset of $Q$).
A {\em run} of $\ar$ on a $\S$-labeled tree $\lambda$ is a labeling 
$\rho:\{a,b\}^*\vers Q$ such that
\debeqno
	\rho(\eps)\in I\mbox{ and }\forall u\in T:
		(\rho(u),\lambda(u),\rho(ua),\rho(ub))\in \D \ .
\fineqno
A run is accepting iff on each (infinite) branch of the run, the set of
states appearing infinitely often is equal to one of the $F\in\fr$.
A tree is accepted iff there exists an accepting run.
Muller automata cannot be determinized in general, but the class of languages
accepted by Muller automata is closed under union, intersection, projection and
complementation.
In particular an automaton accepting all trees where only one node is labeled by $1$
cannot be deterministic: it has to guess where is the $1$.

Formulas of the {\em (full) Monadic Second Order Logic} (MSO) over $(\w^\w,<)$ are built from 
\debitem
\item a countable set of first order variables $x,y,z,\dots$
\item a countable set of second order variables (in capitals) $X,Y,Z,\dots$
\item the order relation $(x<y)$ over first order variables,
\item the membership relation $(x\in X)$, also written $X(x)$,
\item the Boolean connectives $\neg$, $\wedge$ and $\vee$ ($\imp$ and $\equ$ are
seen here as abbreviations),
\item existential quantification $(\exists)$ over first order {\em and} second order 
variables ($\forall$ is seen as an abbreviation of $\neg\exists\neg$).
\finitem
The syntax of the {\em Weak Monadic Second Order Logic} (WMSO) is exactly the
same, the difference is that second order variables are interpreted by {\em finite} subsets of the structure.
\debexm
The formulas of Example~\ref{exm:1} above are also expressible in $\MSO(\w^\w,<)$
because they do not need the addition. One can also define a relation $x=y+1$.
The next formula shows that the set of even ordinals (less than $\w^\w$) can
be defined in MSO:
\debeqno
	\exists X: \forall x & & (x\in X \equ \neg(x+1\in X))\wedge
				(\neg\exists y (x=y+1)\imp x\in X)\ .
\fineqno	
\finexm
We consider trees labeled over $\{0,1\}^k$ where $k$ is the number of first-order
{\em and} second-order free variables.
It should be clear that one can construct Muller automata recognizing
the relations $x\in X$ and $x<y$. Note that for each first-order variable the
automaton has to check that only one node in the tree is labeled by $1$,
\ie, $x$ is treated as a second-order variable $X=\{x\}$.
See \cite{Bedon01} for a clear exposition of a similar construction in the
framework of ordinal automata.
\debthm \label{thm:MSO1}
The Monadic Second Order Theory of $(\w^2,<)$ is decidable in time $\bigO(\tow(n,c))$, 
for some constant $c$, where $n$ is the length of the formula.
\finthm
Recall that the upper bound of \cite{Maurin96} is in $\bigO(\tow(6n,1))$ for
the {\em weak} variant $\WMSO(\w^\w,<)$. Already $\MSO(\w,<)$ has a lower
bound in $\Omega(\tow(n,d))$ for some constant $d>0$ \cite{Reinhardt01}, 
so our bound is really tight.
\debdem[sketch]
We use again the well known method by induction on the structure of the formula
$\p\in \MSO(\w^2,+)$. 
\debitem
\item If $\p$ is an atomic proposition, it is clear how to construct $\ar_\p$.
\item If $\p$ is of the form $\neg \p'$, $\p_1\vee \p_2$ or $\p_1\wedge \p_2$, 
      we use the fact that languages of Muller tree automata are closed under
      complementation, union and intersection.
\item If $\p$ is of the form $\exists x \p'$ or $\exists X \p'$, we use the fact
      that languages of Muller tree automata are closed under projection.
\finitem
The most expensive step is the complementation, it can be done in
exponential time, and the result has also exponential size,
 see \cite{Niessner01,Weyer01}. 
At the end the test of emptiness is also exponential.
\findem 
Note that for the case of disjunction the automaton has to guess at the root
which subformula can be true. For a formula $\p=\p_1\imp \p_2$ we cannot do
better than transform it into $\neg\p_1 \vee \p_2$. It is not correct to simply
build the product of $\ar_{\p_1}$ and $\ar_{\p_2}$ and adapt the acceptance
component, because the acceptance condition is checked independently on each 
branch.

Using an idea similar to that of \Sec{sec:beyond}, one can attach $\w$ trees  of
the form presented above to a left-most branch to encode subsets of $\w^3$.
This can be extended by induction to $\w^i$ for all $i<\w$. 
\debthm \label{thm:MSO2}
For each $i<\w$ there exists a constant $c_i$ such that the 
Monadic Second Order Theory of $(\w^i,<)$ is decidable in time 
$\bigO(\tow(n,c_i))$, where $n$ is the length of the formula.
\finthm
In other works such as \cite{CartonRispal04,Bedon96} the emphasis is not placed on
the complexity, but it seems that the complementation of ordinal automata is
double exponential.
It is open how to extend the tree encoding to subsets of $\w^\w$.

\subsection{MSO-interpretation. Comparison with Ordinal Automata}
%

It is possible to put a different light on the previous constructions. 
The MSO theory of the full binary tree \cite{Weyer01}, 
called {\em S2S}, is build from the 
atomic propositions $S_a(x,y)$, $S_b(x,y)$ and $P(x)$, where $S_a$ is the
relation ``left successor'', $S_b$ is ``right successor'' and $P$ is
a predicate that indicates that the label of a node is $1$. In other words,
given a labeled infinite tree $\lambda:\{a,b\}^*\vers \{0,1\}$ and
$x,y \in \{a,b\}^*$:
\debeqno
        S_a(x,y)\Equ y=x.a\ , \qquad S_b(x,y) \Equ y=x.b\ , \qquad 
	                P(x)  \Equ \lambda(x)=1\ .
\fineqno
Recalling the left figure in page~\pageref{fig:positions},
the order among the ordinals/positions in the tree can be interpreted in S2S.
That is, one can write a formula $\phi(x,y)$ such that $\phi(x,y)$ is true iff
the ordinal of position $x$ is less than that of $y$. It is easy if one first 
write formulas $\phi_a(x,y)$ and $\phi_b(x,y)$ that checks that $y$ is a left 
descendant of $x$ (resp. right descendant). 
\\
\begin{minipage}[b]{6.5 cm}
Alternatively one can see the ordering
$\w^2$ as the transitive closure of the graph pictured on the right.
Nevertheless concerning complexity it
is better to construct dedicated automata as in the proof of \Thm{thm:MSO1}.
In other words the graphs of the orderings $\w^i$, $i<\w$, are prefix-recognizable graphs \cite{Caucal02}. It is open whether graphs of greater ordinals are in the Caucal
hierarchy.
\end{minipage}
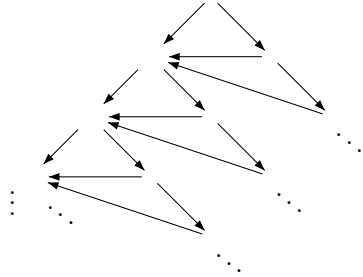
\hf
{ \unitlength=0.8mm
\begin{picture}(60,48)(-34,-47)
        \gasset{Nadjust=wh,Nadjustdist=1.5,Nframe=n}
	\node(r)(0,0){$$}
	\node(1)(10,-10){$$}
	\node(11)(20,-20){$$}
	\node(111)(23,-23){$\ddots$}		
	\node[Nadjustdist=3](0)(-10,-10){$$}
	\node(01)(0,-20){$$}
	\node(011)(10,-30){$$}
	\node(0111)(13,-33){$\ddots$}
	\node[Nadjustdist=3](00)(-20,-20){$$}
	\node(001)(-10,-30){$$}
	\node(0011)(0,-40){$$}
	\node(00111)(+3,-43){$\ddots$}
	\node[Nadjustdist=3](000)(-30,-30){$$}
	\node(0001)(-25,-35){$\ddots$}
	\node(0000)(-33,-33){$\vdots$}

	\gasset{AHnb=1}
	\drawedge(r,1){$$}
	\drawedge(1,11){$$}
	\drawedge(r,0){$$}
	\drawedge(1,0){$$}
	\drawedge(11,0){$$}
	\drawedge(0,01){$$}
	\drawedge(01,011){$$}
	\drawedge(0,00){$$}
	\drawedge(01,00){$$}
	\drawedge(011,00){$$}
	\drawedge(00,001){$$}
	\drawedge(001,0011){$$}
	\drawedge(00,000){$$}
	\drawedge(001,000){$$}
	\drawedge(0011,000){$$}
\end{picture} }

The usual proof that $\MSO(\w^\w,<)$ is decidable uses ordinal automata
reading ordinal words. An ordinal word of length $\a$ is a mapping
$\a \vers \S$, where $\S$ is a finite alphabet. An ordinal automaton has
a state space $Q$, usual one-step transitions of the form 
$(q,\s,q')\in Q\ti\S\ti Q$
and {\em limit transitions} of the form $(P,q')\in \pr(Q)\ti Q$, 
see e.g. \cite{Bedon01}. They are a generalization of Muller (word)
automata. A run is a mapping $\rho:\a+1\vers Q$. For a successor ordinal
$\b+1$, $\rho(\b+1)$ is defined in the usual way. For a limit ordinal
$\b$, the state $\rho(\b)$ is obtained by a limit transition according
to the states appearing infinitely often ``before'' $\b$.

We want to point out that a run of a Muller automaton on a tree 
representing $S\inc \w^2$ is very similar to a run of length $\w^2$
of an ordinal automaton.
Consider a node $v$ at depth $i$ on the left most branch. It corresponds to an 
ordinal $\w.i$. The  right-most branch from $v$ must satisfy
the Muller condition, and the state reached at the left successor of $v$ 
is like the state reached at the limit transition at $\w.(i+1)$.
In this way we get a new proof that languages accepted by ordinal automata
are closed under complementation, restricted to the case of words of length
$\w^j$, for all $j<\w$.

Comparing both approaches, we see that tree automata can not be determinized
in general, they can be complemented, however, using an exponential construction.
On the other side ordinal automata can be determinized (and complemented)
using a doubly exponential construction, due to the nesting of Muller conditions.
We are not aware of a better complementation algorithm for ordinal automata,
see e.g. \cite{CartonRispal04} for a more general result.
The transformation from a tree automaton to an equivalent ordinal automaton
according to our coding is very simple. The state space remains the same
except for one extra final state for the last limit transition. If $(q,\lambda,q_a,q_b)\in \D$
in the tree automaton, add transitions $(q,\lambda,q_b)$, and $(P,q_a)$ for
all $P\in\fr$, where $\fr$ is the Muller acceptance condition. The other way around
is more complicated because the tree automaton has to guess what states are
going to be visited infinitely often on the right branch, and then allow
only these states to be visited infinitely often.

\subsection{Weak MSO and FO} \label{sub:WMSO-FO}
%

We introduce here new material to compare MSO and FO.
Any ordinal $\b$ can be written in a unique way in the form 
\debeqno
       2^{\g_{n-1}}+\dots+2^{\g_0}\ ,\ \mbox{ where }\ \ (\g_{n-1},\dots,\g_0)
\fineqno
is a strictly decreasing sequence of ordinals. The set 
$\{\g_{n-1},\dots,\g_0\}$ is called the {\em 2-development} of $\b$.
For example $2^\w=\w$, $2^{\w.i+j}=2^{\w.i}.2^j=\w^i.2^j$, 
$2^{\w^2}=\left(2^\w\right)^\w=\w^\w$. Let $E$ be
the binary relation on ordinals such that $(x,y)\in E$ iff $x=2^\g$ for
some $\g$ that belongs to the 2-development of $y$.
It is known \cite{Buechi65a} that the theories $\WMSO(\a,<)$
and $\FO(2^\a,+,E)$ are equireducible in linear time. Recall that the (weak)
theory WMSO is the monadic theory where only finite sets are considered.
This mean that any formula of one of the logics can be translated into an
equivalent formula of the other logic in linear time.

To extend \Thm{thm:FO2} to the decidability of $\FO(2^\a,+,E)$ for $\a=\w^i$, we only need a tree automaton recognizing the relation $E$. The fact that $x=2^\g$ is
equivalent in our coding to the fact that exactly one label is $1$ in the
tree $T_x$, and $(x,y)\in E$ if moreover the same node is
labeled by $1$ in the tree $T_y$. The automaton recognizing 
$E$ needs only three states, so the complexity bounds of \Thm{thm:FO2}
are not changed.

On the other side we have proved decidability of the {\em full} MSO
theory of $(\w^i,<)$ in \Thm{thm:MSO2}. 
It remains to interpret WMSO in MSO. 
It is known in general how to construct a Muller tree automaton that checks 
that only finitely many nodes of a tree are labeled by $1$. It is possible with 
only $2$ states and can be used to adapt the proof of Theorem~\ref{thm:MSO1}
to WMSO.
Using this reduction, the complexity
of the decision procedure of $\WMSO(\w^i,<)$ is in $\bigO(\tow(n+1,c'_i))$
for some (new) constant $c'_i$. Alternatively, using the property that every subset 
of an ordinal is also well ordered, it is possible to write an MSO 
formula that checks that a set of ordinals is finite. This formula should be used
together with each second order quantification.

An extension of the previous tree-automata techniques to higher ordinals such as
$\MSO(\w^\w,<)$ would gives also tree-automata techniques for $\WMSO(\w^\w,<)$
and then $\FO(\w^{\w^\w},+,E)$, which is impossible \cite{Delhomme04} 
(see end of Section~\ref{sec:FO}).

Related to the Cantor Normal Form (see \Sec{sec:FO}), {\em any} ordinal $\b$ 
can yet be written in a unique way in the form 
\debeqno
	\a=\g.\w^\w+\w^p n_p+\w^{p-1}n_{p-1}+\dots+  \w^{1}n_1+n_0 \ .
\fineqno
where $n_p>0$. The $\w$-character of $\a$ is the sequence 
$(\sigma, n_p,\dots,n_0)$ where $\sigma=0$ if $\g=0$, and $\sigma=1$ if 
$\g>0$. The theories $WMSO(\a,<)$ and $WMSO(\b,<)$ are equal iff $\a$
and $\b$ have the same $\w$-character \cite{Buechi65a}. It follows that
$\FO(2^\a,+,E)$ and $\FO(2^\b,+,E)$ are equal iff $\a$ and $\b$ have the 
same $\w$-character.

%
%
\section{Perspectives}
%

We gave a new decision procedure for $\FO(\w^{\w^i},+)$ and $\MSO(\w^i,<)$
achieving better complexity bounds.
We hope our constructions are easy to understand. As a byproduct we have
a new proof of the complementation of ordinal automata restricted to words
of length $\w^i$.

According to \cite{Delhomme04} (see end of Section~\ref{sec:FO})
and Section~\ref{sub:WMSO-FO}  it is not 
possible to extend the tree-automata techniques to higher ordinals.
But we would like to extend it to other linear orderings.
A bi-infinite word is a mapping from the {\em relative} integers to a finite 
alphabet. It is easy to represent it as an infinite tree where only the right
most and the left most branches are relevant. It seems easy to represent also 
orderings like $-\w$ or $\w\times(-\w)$. Using a special letter, one could mark
branches where the ``reverse'' ordering $-\-w$ is used.
We conjecture that one can extend the
results of Section~\ref{sec:MSO} to more general linear orderings than just ordinals,
and give a new proof of the results of \cite{CartonRispal04}.

%
%
\subsection*{Acknowledgments}
%

Many thanks to Wolfgang Thomas for always saying that we have a proof 
without theorem (the proof that tree automata are closed under intersection,
complementation and projection), to him and Stéphane Demri for pointing out 
some useful references
and to the referees.

\bibliographystyle{plain}
\bibliography{ordinalsFO}

\end{document}

%% file: macros-ordinalsFO.tex
\usepackage{mathrsfs}
\usepackage[latin1]{inputenc}

\newcommand{\ie}{{\it i.e.}}

\newcommand{\hf}{\hfill}

\newcommand{\vers}{ \mapsto}
\newcommand{\tr}[3]{\mbox{ \scriptsize $\begin{array}{c} #1\\ #2\\ #3 \end{array}$ }}

\newcommand{\FO}{\mbox{FO}}
\newcommand{\MSO}{\mbox{MSO}}
\newcommand{\WMSO}{\mbox{WMSO}}
\newcommand{\tow}{\mbox{Tower}}

\newcommand{\Thm}[1]{Theorem~\ref{#1}}
\newcommand{\Sec}[1]{Section~\ref{#1}}

\newcommand{\debcor}{\begin{corollary} }
\newcommand{\debexm}{\begin{example} }
\newcommand{\debdem}{\begin{proof} }
\newcommand{\debctr}{\begin{center} }
\newcommand{\debthm}{\begin{theorem} }
\newcommand{\debarr}{\begin{array} }
\newcommand{\debitem}{\begin{itemize} }
\newcommand{\debeqn}{\begin{eqnarray}}   
\newcommand{\debeqno}{\begin{eqnarray*}} 

\newcommand{\fincor}{\end{corollary} }
\newcommand{\finexm}{\end{example} }
\newcommand{\findem}{\end{proof} }
\newcommand{\finctr}{\end{center} }
\newcommand{\finthm}{\end{theorem} }
\newcommand{\finarr}{\end{array} }
\newcommand{\finitem}{\end{itemize} }
\newcommand{\fineqn}{\end{eqnarray}}   
\newcommand{\fineqno}{\end{eqnarray*}} 

\newcommand{\imp}{ \rightarrow}
\newcommand{\equ}{ \leftrightarrow}
\newcommand{\Imp}{ \Rightarrow}
\newcommand{\Equ}{ \Leftrightarrow}

\newcommand{\ti}{\times}

\newcommand{\inc}{\subseteq}
\newcommand{\lc}{\left<}
\newcommand{\rc}{\right>}

\newcommand{\p}{\psi}
\renewcommand{\a}{\alpha}
\renewcommand{\b}{\beta}
\newcommand{\g}{\gamma}
\renewcommand{\d}{\delta}
\newcommand{\eps}{\varepsilon}
\newcommand{\s}{\sigma}
\newcommand{\w}{\omega}
\newcommand{\D}{\Delta}
\renewcommand{\S}{\Sigma}

\newcommand{\ar}{ \mathcal{A}} 
\newcommand{\fr}{ \mathcal{F}} 

\newcommand{\pr}{\mathscr{P}} 

\newcommand{\bigO}{ \mathcal{O}}




%% file: ordinalsFO.bbl
\begin{thebibliography}{10}

\bibitem{Bedon96}
Nicolas Bedon.
\newblock Finite automata and ordinals.
\newblock {\em Theor. Comput. Sci.}, 156(1{\&}2):119--144, 1996.

\bibitem{Bedon01}
Nicolas Bedon.
\newblock Logic over words on denumerable ordinals.
\newblock {\em J. Comput. System Sci.}, 63(3):394--431, 2001.

\bibitem{BruyCartonSeniz03}
V{\'e}ronique Bruy{\`e}re, Olivier Carton, and G{\'e}raud S{\'e}nizergues.
\newblock Tree automata and automata on linear orderings.
\newblock In {\em Proceedings of WORDS'03}, volume~27 of {\em TUCS Gen. Publ.},
  pages 222--231. Turku Cent. Comput. Sci., Turku, 2003.

\bibitem{Buechi60}
J.~Richard B{\"u}chi.
\newblock Weak second-order arithmetic and finite automata.
\newblock {\em Zeitschrift f{\"u}r mathematische Logik und Grundlagen der
  Mathematik}, 6:66--92, 1960.

\bibitem{Buechi62}
J.~Richard B{\"u}chi.
\newblock On a decision method in restricted second order arithmetic.
\newblock In {\em Proceedings of the 1960 International Congress on Logic,
  Methodology and Philosophy of Science}, pages 1--11. Stanford University
  Press, 1962.

\bibitem{Buechi65a}
J.~Richard B{\"u}chi.
\newblock Decision methods in the theory of ordinals.
\newblock {\em Bull. Amer. Math. Soc.}, 71:767--770, 1965.

\bibitem{icalp03}
Thierry Cachat.
\newblock Higher order pushdown automata, the {C}aucal hierarchy of graphs and
  parity games.
\newblock In {\em Proceedings of the 30th International Colloquium on Automata,
  Languages, and Programming, ICALP'03}, volume 2719 of {\em LNCS}, pages
  556--569. Springer, 2003.

\bibitem{CartonRispal04}
Olivier Carton and Chloe Rispal.
\newblock Complementation of rational sets on scattered linear orderings of
  finite rank.
\newblock In Martin Farach-Colton, editor, {\em LATIN}, volume 2976 of {\em
  LNCS}, pages 292--301. Springer, 2004.

\bibitem{Caucal02}
Didier Caucal.
\newblock On infinite terms having a decidable monadic theory.
\newblock In {\em Proceedings of the 27th International Symposium on
  Mathematical Foundations of Computer Science 2002, MFCS 2002}, volume 2420 of
  {\em LNCS}, pages 165--176. Springer, 2002.

\bibitem{tata97}
Hubert Comon, Max Dauchet, Remi Gilleron, Florent Jacquemard, Denis Lugiez,
  Sophie Tison, and Marc Tommasi.
\newblock Tree automata techniques and applications.
\newblock Available on: \url{http://www.grappa.univ-lille3.fr/tata}, 1997.
\newblock release October, 1rst 2002.

\bibitem{Delhomme04}
Christian Delhomm{\'e}.
\newblock Automaticit\'e des ordinaux et des graphes homog\`enes.
\newblock {\em C. R. Math. Acad. Sci. Paris}, 339(1):5--10, 2004.

\bibitem{Dershowitz93}
Nachum Dershowitz.
\newblock Trees, ordinals and termination.
\newblock In Marie-Claude Gaudel and Jean-Pierre Jouannaud, editors, {\em
  TAPSOFT}, volume 668 of {\em LNCS}, pages 243--250. Springer, 1993.

\bibitem{dagst}
Erich Gr{\"a}del, Wolfgang Thomas, and Thomas Wilke, editors.
\newblock {\em Automata, Logics, and Infinite Games: A Guide to Current
  Research}, volume 2500 of {\em LNCS}. Springer, 2002.

\bibitem{Klaedtke04}
Felix Klaedtke.
\newblock On the automata size for {Presburger} arithmetic.
\newblock In {\em Proceedings of the 19th Annual IEEE Symposium on Logic in
  Computer Science (LICS 2004)}, pages 110--119. IEEE Computer Society Press,
  2004.
\newblock A full version of the paper is available from the author's web page.

\bibitem{KupfermanVardi00}
Orna Kupferman and Moshe~Y. Vardi.
\newblock An automata-theoretic approach to reasoning about infinite-state
  systems.
\newblock In E.~A. Emerson and A.~P. Sistla, editors, {\em Proceedings of the
  12th International Conference on Computer Aided Verification, CAV'00}, volume
  1855 of {\em LNCS}, pages 36--52. Springer, 2000.

\bibitem{Maurin96}
Fran\c{c}oise Maurin.
\newblock Exact complexity bounds for ordinal addition.
\newblock {\em Theoretical Computer Science}, 165(2):247--273, 1996.

\bibitem{Niessner01}
Frank Nie{\ss}ner.
\newblock Nondeterministic tree automata.
\newblock In Gr{\"a}del et~al. \cite{dagst}, pages 135--152.

\bibitem{Rabin69}
Michael~O. Rabin.
\newblock Decidability of second-order theories and automata on infinite trees.
\newblock {\em Transactions of the American Mathematical Society}, 141:1--35,
  1969.

\bibitem{Reinhardt01}
Klaus Reinhardt.
\newblock The complexity of translating logic to finite automata.
\newblock In Gr{\"a}del et~al. \cite{dagst}, pages 231--238.

\bibitem{Rosenstein82}
Joseph~G. Rosenstein.
\newblock {\em Linear orderings}.
\newblock Academic Press Inc. [Harcourt Brace Jovanovich Publishers], New York,
  1982.

\bibitem{Sierpinski65}
Wac{\l}aw Sierpi{\'n}ski.
\newblock {\em Cardinal and ordinal numbers}.
\newblock Second revised edition. Monografie Matematyczne, Vol. 34. Pa\'nstowe
  Wydawnictwo Naukowe, Warsaw, 1965.

\bibitem{Weyer01}
Mark Weyer.
\newblock Decidability of {S1S} and {S2S}.
\newblock In Gr{\"a}del et~al. \cite{dagst}, pages 207--230.

\end{thebibliography}
